\documentclass[a4paper,10pt,twocolumn]{article}
\usepackage{graphicx}
\usepackage[margin=0.5in]{geometry}

\begin{document}

\title{Ultra low-loss, isotropic 2D optical negative-index metamaterial based on hybrid metal-semiconductor nanowires}
\author{R. Paniagua-Dom\'inguez, D. R. Abujetas and J. A. S\'anchez-Gil\\\\
Instituto de Estructura de la Materia, Consejo Superior de Investigaciones Cient\'{\i}ficas\\ Serrano 121, 28006 Madrid, Spain}

\maketitle

\begin{abstract}
In the past few years, many of the fascinating, and previously almost unconceivable, properties predicted for those novel, artificial, man-made materials, so called metamaterials, were demonstrated to be not only a tangible reality, but a very useful one. However, plenty of the best achievements in that newly discovered field are far from having a direct translation to the, in many aspects more interesting, high frequency range, without being
burdened not only by technological difficulties, but also conceptual ones. Of particular importance within the realm of optical metamaterials 
having a negative index of refraction, often designated negative-index metamaterials (NIM),  
is the issue of simultaneously achieving a strong response of the system and low associated losses. In the present work, we demonstrate the possibility to use hybrid metal-semiconductor nanowires to obtain an isotropic optical NIM,
with very strong electric and magnetic responses, which exhibits
extremely low losses  (about two orders of magnitude better than present optical NIMs).  

Artificial magnetism of the effective medium is achieved by exciting the lowest order Mie-like 
magnetic resonance in the wires, while we show that it is possible to
tune the geometrical parameters of the system to make this resonance spectrally overlap
with the lowest electric, dipole-like, one. For such a goal, we propose to use noble metals 
such as silver or gold to build the core, and a high permittivity semiconductor, such as silicon or germanium, to build the 
shell. Polarization of the incident wave with the magnetic field along the axis of the wires will generate the desired 
responses. The metamaterial obtained with these building blocks is shown to produce negative refraction in the near-infrared, with values
of the real part of the index well below -1, and figures of merit extraordinarily high.  Although shown here at $\sim1.3-1.5\,\mu$m, the
tunability of the system allows to select the operating range in the whole telecom frequency spectrum. The 
proposed design is tested to work in several interesting configurations, including prisms and slabs, in which Snell's law allows one
to directly observe the negative refraction, and with different spatial arrangements.
\end{abstract}

The so called metamaterials are artificial materials in which the effective medium properties, usually
exotic and not naturally attainable, depend on the geometry of their basic constituents, rather than on their chemical composition. 
Ranging from sub-diffraction resolution \cite{PendryPRL,NatMat} to spontaneous emission  \cite{HyperMet} to extreme control over the flow of light \cite{TO,TO2}, many exciting 
and unexpected phenomena have been achieved or predicted for this new kind of materials. Although originally developed in 
electromagnetics, many of the ideas developed in this field have been successively extended or adapted to other ondulatory phenomena, such
as acoustics, making it one of the most active in the engineering and physical sciences in the past few years.

Still, however, there are many open challenges in this field. Among them, the realization of a bulk isotropic negative index metamaterial (NIM), 
with negative refraction and low losses in the optical domain. In general terms, the major issue when trying to achieve such a goal
is obtaining a strong diamagnetic response of the constituents, enough to lead to an effective negative permeability. Probably 
as a consequence of the success of the original designs operating in the microwave regime, many efforts were made to adapt
them to increasingly higher frequencies, mainly by miniaturization \cite{OptMet,WegSou}. Apart from some inherited from the original designs,
such as anisotropy, many drawbacks were found in doing so, mainly related to the different behavior of  metals at optical 
frequencies, such as saturation of the magnetic response \cite{PRLSat} or high losses associated to ohmic currents, with maximum figures of merit (f.o.m$=-Re(n_{eff})/Im(n_{eff})$) of up to f.o.m.$\sim 3$ \cite{PRLfish}. As a consequence, 
completely different strategies were studied to obtain artificial magnetism, as those based in displacement currents appearing in nanoparticle 
clusters due to coupling between structures \cite{Engheta}. However, some of the most 
successful were those which attempted to obtain it from natural magnetic resonant modes in high permittivity 
dielectrics, leading to low-loss magnetic materials \cite{Marques,Wheeler,PRLPopa,G-E}. Secondary structures or particular
arrangements were needed to provide the additional electrical response, necessary to obtain doubly negative index of refraction
\cite{MatTod,PRLDie,PRLDie2}.

In this work, we propose a structure that, combining electric and magnetic responses, can be used as the
basic building block for extremely low-loss (f.o.m.$ \sim 300$, isotropic two-dimensional metamaterials in the near-infrared, with 
simultaneously negative permittivity ($\epsilon$) and permeability ($\mu$) at optical frequencies. Such structure is a metallo-dielectric 
core-shell nanowire of circular cross section. The electric response being due to the excitation of a localized surface plasmon resonance (LSPR) 
in the metallic core, requires the polarization to be fixed with the magnetic field along the axis of the nanowire (TE-polarization).

The optical properties of coated cylinders under plane wave illumination can, indeed, be analytically studied \cite{Kerker}. Under 
TE polarized light, the scattering and extinction efficiencies of these particles can be written as sums over different coefficients:
\begin{equation}
 Q_{sca}=\frac{2}{x}(|a_0|^2+2\sum_{j=1}^\infty |a_j|^2)
\end{equation}
\begin{equation}
Q_{ext}=\frac{2}{x}Re(a_0+2\sum_{j=1}^\infty a_j )
\end{equation}
which can be then assigned a certain multipolar character, either magnetic or electric. 
Although the expressions are similar to those of a solid cylinder (i.e., without coating), 
the $a_j$ coefficients contain information about both the core and the shell. 
With this given polarization, the first
zero-order coefficient, $a_0$, has magnetic character, while the second coefficient, $a_1$, can be identified as the electric
dipolar contribution in the 2D plane perpendicular to the cylinder axis.
\begin{figure}
\includegraphics[width=0.9\columnwidth]{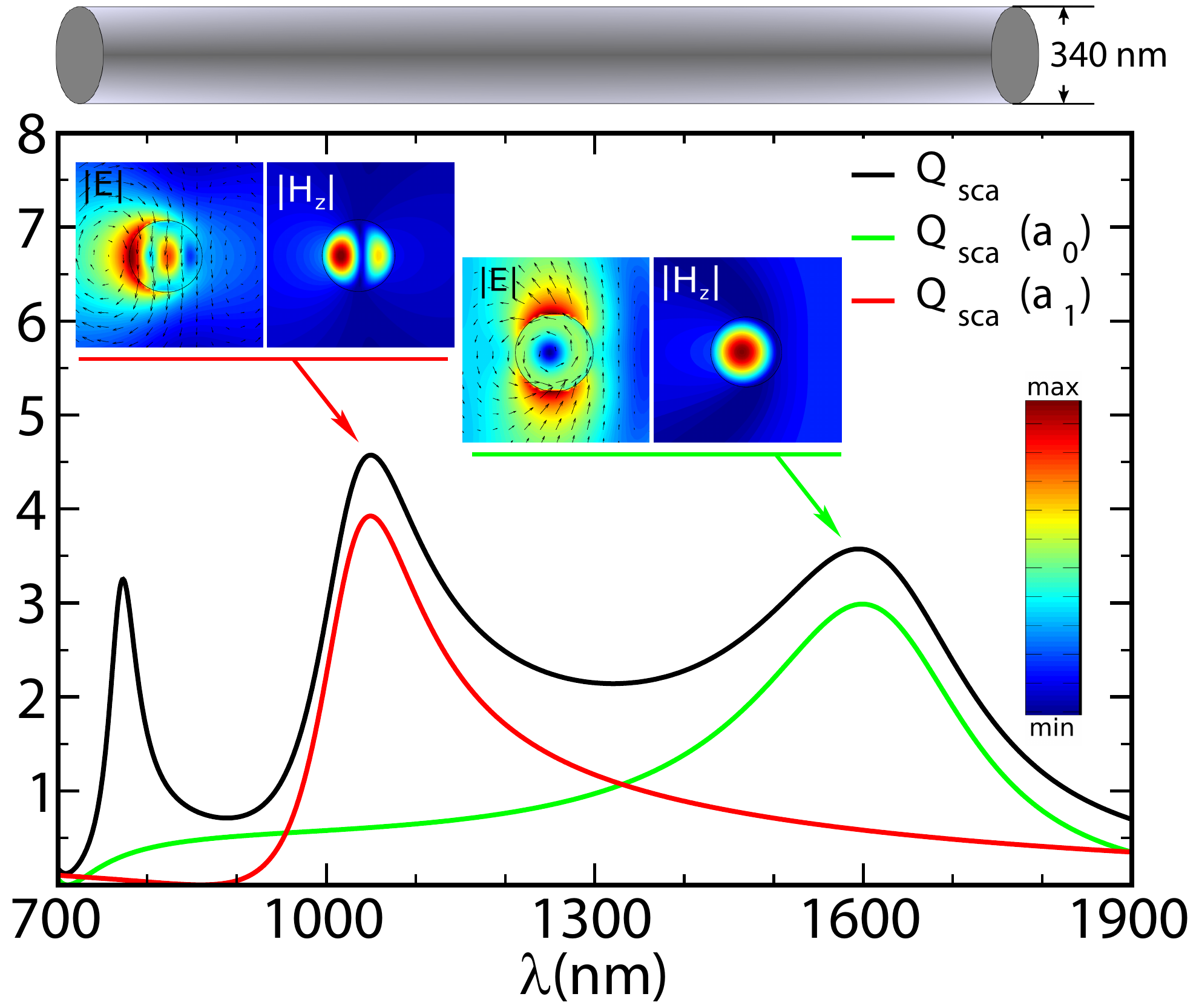}
\caption{Scattering efficiency for a Si cylinder with radius $R=170$~nm. The contributions of the two first terms in the
expansion of the field are also shown. Depicted in the inset are the maps of the near electric field (left), and out-of-plane
component of the magnetic field (right) at the magnetic (green, $\lambda_0=1605$~nm) and electric (red, $\lambda_0=1050$~nm) resonances. 
The TE-polarized plane wave impinges from the right hand side.}
\label{figure1}
\end{figure}
We can realize that this is the case if we study
the optical properties of a pure silicon (Si) cylinder. Incidentally, Silicon spheres \cite{G-E} have attracted the 
interest of the community, because of their potential use in metamaterials and interesting radiative properties. In 
Fig.\ref{figure1} we have plotted the scattering efficiency of a solid Si cylinder of radius $R=170$~nm under normal incidence of a
TE-polarized plane wave. It can be immediately realized that the optical properties are quite similar to that of a pure Si sphere \cite{G-E}, although the correspondence between Mie coefficients and electric and magnetic multipolar terms is not straightforward for cylinders. Thus, we need to identify the character of the $a_0$ and $a_1$ coefficients. To this end, we plot the total electric field and the out-of-plane component (only non-zero) 
of the magnetic field In the insets of Fig.\ref{figure1}  for the first two resonances. Similar to the case of a sphere, the magnetic resonance is the one excited at lower frequencies (green curve), thus corresponding to the $a_0$ coefficient.
 The strong circulation of the electric field is a clear signature of the
magnetic character of the resonance. On the other hand, the subsequent higher-energy 
resonance reveals the dipolar electric character of the $a_1$ resonance.

\begin{figure}
\includegraphics[width=\columnwidth]{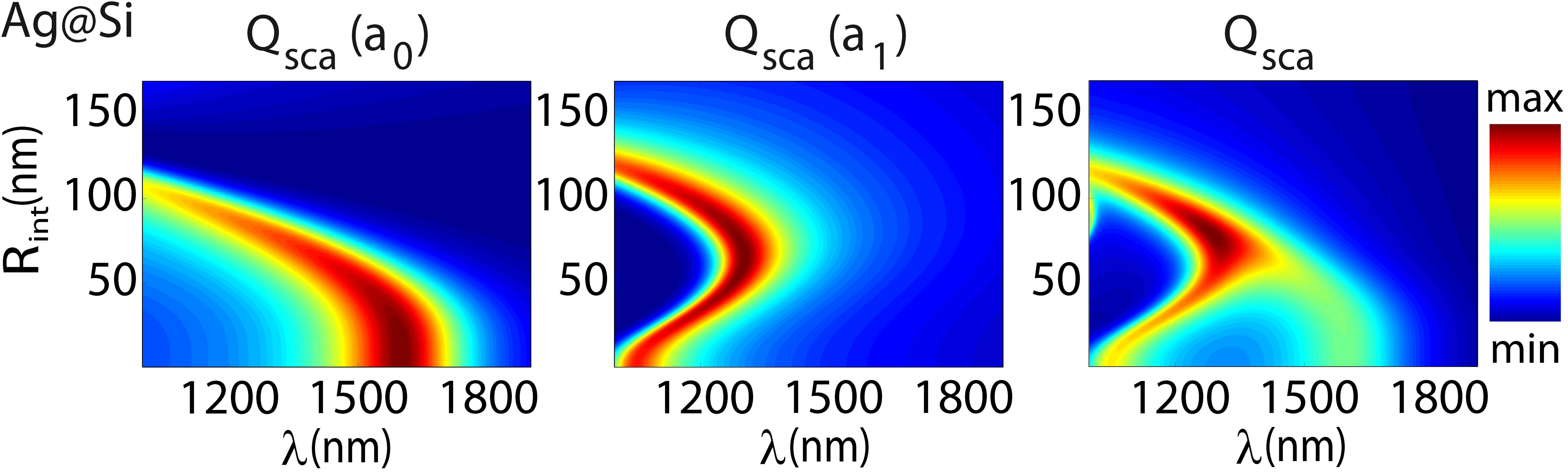}
\caption{Magnetic (left) and dipolar electric (center) contributions to the total scattering efficiency (right) of a 170~nm Ag covered with Si (Ag@Si) core-shell nanocylinder.}
\label{figure2}
\end{figure}

\begin{figure}
\includegraphics[width=\columnwidth]{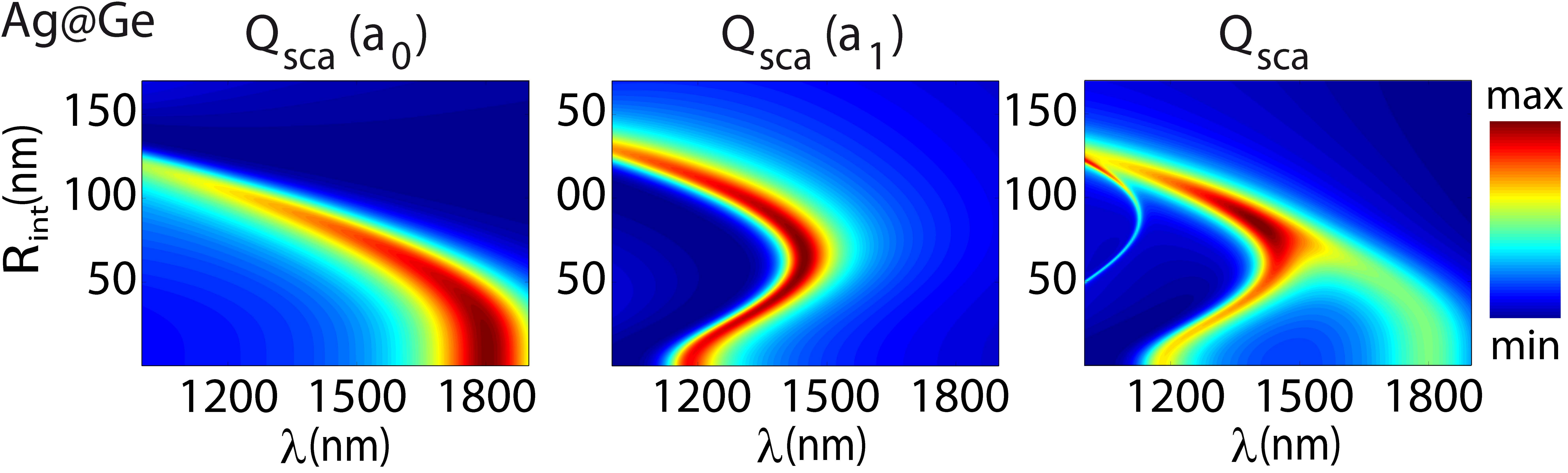}
\caption{Magnetic (left) and dipolar electric (center) contributions to the total scattering efficiency (right) of a 170~nm Ag@Ge core-shell nanocylinder.}
\label{figure3}
\end{figure}

Combining a dielectric shell, which exhibits magnetic properties similar to those of solid cylinders, with a plasmonic core, 
will provide the additional electric resonance necessary to obtain
a doubly resonant structure suitable as a NIM building block (meta-atom). In order to make both resonances spectrally overlap, tuning of the
geometrical parameters of the structure is needed.  This approach has been  successful employed in the case of a core-shell sphere \cite{NJP}, but it has to be proven for a core-shell cylinder due to the different behavior of its dipolar magnetic resonance. 
In Figs.\ref{figure2} and \ref{figure3}, we have plotted the magnetic and electric contributions 
to the scattering, together with the total efficiency, for core-shell nanowires of outer radius $R_{out}=170$~nm as a 
function of the wavelength of incidence and inner radius ($R_{int}$). 

The materials used are silver (Ag) for the core, and Si (Fig.\ref{figure2}) and germanium (Ge, Fig.\ref{figure3}) for the shell. In the case
of Si cover, the electric and magnetic responses spectrally overlap over a relatively wide range, $\lambda_0\sim(1100,1500)$~nm, for
$R_{int}$ values between 60 and 100~nm. This makes the core-shell nanowire design particularly robust against possible
fabrication defects. Notice that, in the limit of $R_{in}\rightarrow0$, we recover those values of the pure Si and Ge cylinders.
Although essentially similar, both resonances get redshifted in the case of Ge coating for the same fixed
$R_{out}$ value, due to the slightly higher index of refraction of this material.

\begin{figure}
\includegraphics[width=\columnwidth]{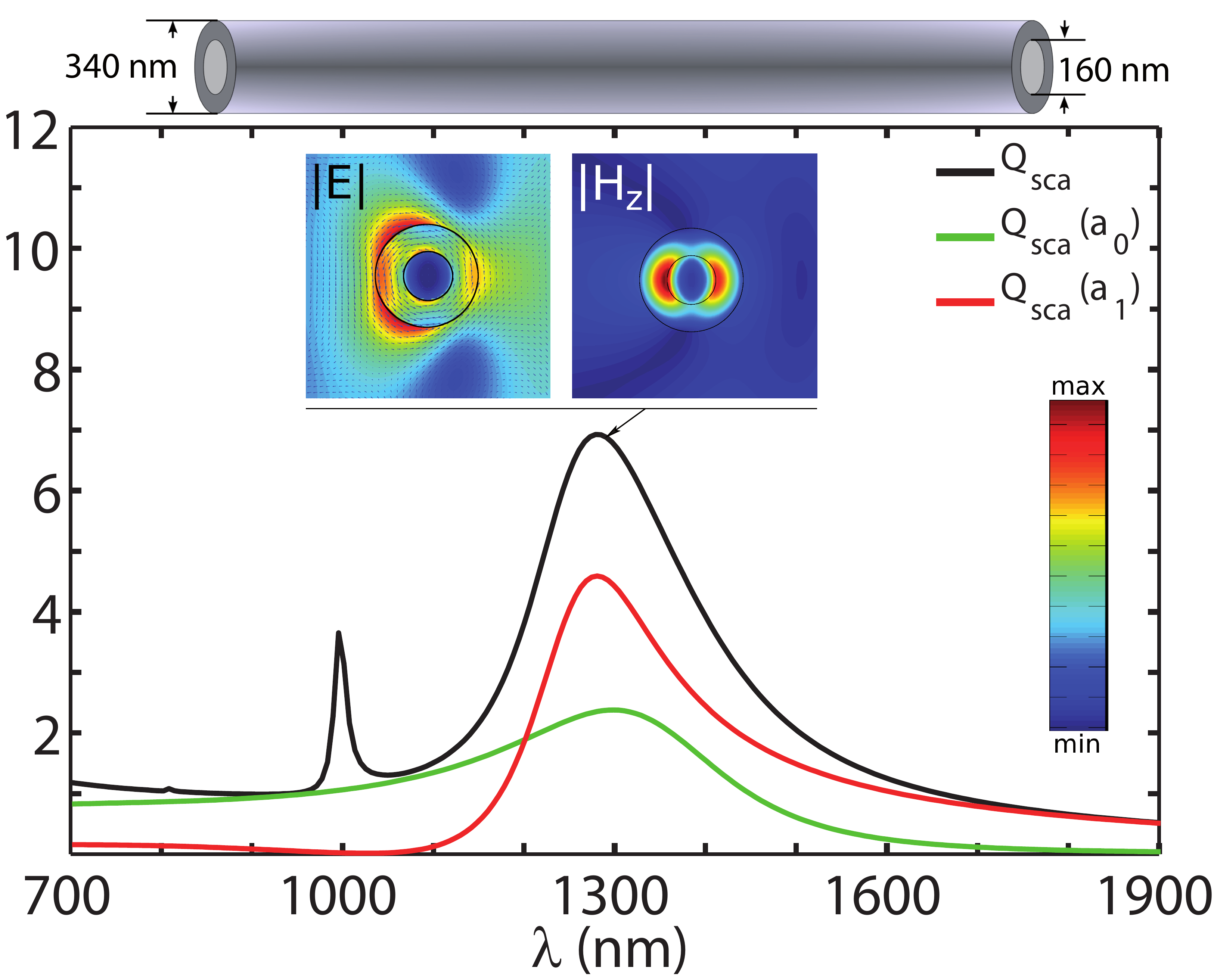}
\caption{Scattering efficiency for a Ag@Si core-shell nanowire with external radius $R_{out}=170$~nm and internal radius $R_{int}=80$~nm.
The contributions of the two first terms in the expansion of the field are also shown. Depicted in the inset are the maps of the near electric field (left), and out-of-plane
component of the magnetic field (right) at resonance, i.e., at $\lambda_0=1298$~nm. The TE-polarized plane wave impinges from the right hand side.}
\label{figure4}
\end{figure}

Let us choose a particular value of $R_{int}$ in the case of, e.g., Ag@Si (totally analogous results hold for Ag@Ge), namely 
$R_{int}=80$~nm. In Fig.\ref{figure4} the total scattering efficiency is plotted, together with the magnetic ($a_0$) and electric
($a_1$) contributions. These two contributions are totally dominant in the frequency range in which both coincide. Moreover,
plotting the near-field distributions of both the electric and magnetic fields at the frequency of coincidence of the 
resonances allows one to identify  the electric and magnetic character of the combined resonance. Indeed, 
a strong rotation of the electric field is observed, together with the signature in the magnetic field of a LSPR 
excitation in the core of the system (see inset in Fig.\ref{figure4}). 


Metallo-dielectric core-shell nanowires are, therefore, doubly resonant structures, having a relatively small 
electrical size. In the Ag@Si case studied, this electrical size reduces to $R_{out}/\lambda_{res}\sim0.13$. These properties allows one to envisage a possible 
application as building blocks for NIM at optical frequencies, as their spherical geometry counterparts \cite{NJP}. Indeed, for compact arrangements, with
lattice periods of about the size of the diameter of the wires, the metamaterials thus obtained may be well 
described as an effective medium (depending also on the absolute values of the effective index of 
refraction, $n_{eff}$, achieved). These compact arrangements can actually
be employed in realistic designs, if we just impose the structures to be unconnected. This is a necessary requirement in order
to keep the magnetic resonance condition of the system unchanged. Coupling between neighboring particles is small,
since most intense fields are localized in the surface of the core (well inside the shell) and they decay quickly in space. Thus, as will be shown later, no 
significant coupling occurs in general, even if we bring the structures very close \cite{NJP}.

Let us now show that it is, actually, possible to obtain a NIM with these nanoparticles. To do so, we will arrange the meta-atoms in two of the most common 
periodic arrangements, namely, in square and hexagonal lattices. We will simulate infinite slabs of different thicknesses to retrieve the effective 
index of refraction from complex reflection and transmission coefficients \cite{NJP,RetPar,RobRetPar}. Real parts of the effective index of refraction will be, moreover, obtained
by simply applying Snell's law. This will be done under several angles of incidence ($\theta_{i}$). The retrieved parameters will then be tested to fulfill 
Snell's law also in prism configurations. In the following, it will be demonstrated that the metamaterials obtained using both Ag@Si and
Ag@Ge core-shell nanowires exhibit not only a negative index of refraction in the optical frequency domain, isotropic in the plane perpendicular to the nanowire axis; moreover, with extremely low losses (f.o.m.$ \sim 300$), about two orders of magnitude better than the best reported configuration \cite{PRLfish}.

\begin{figure}[!ht]
\includegraphics[width=\columnwidth]{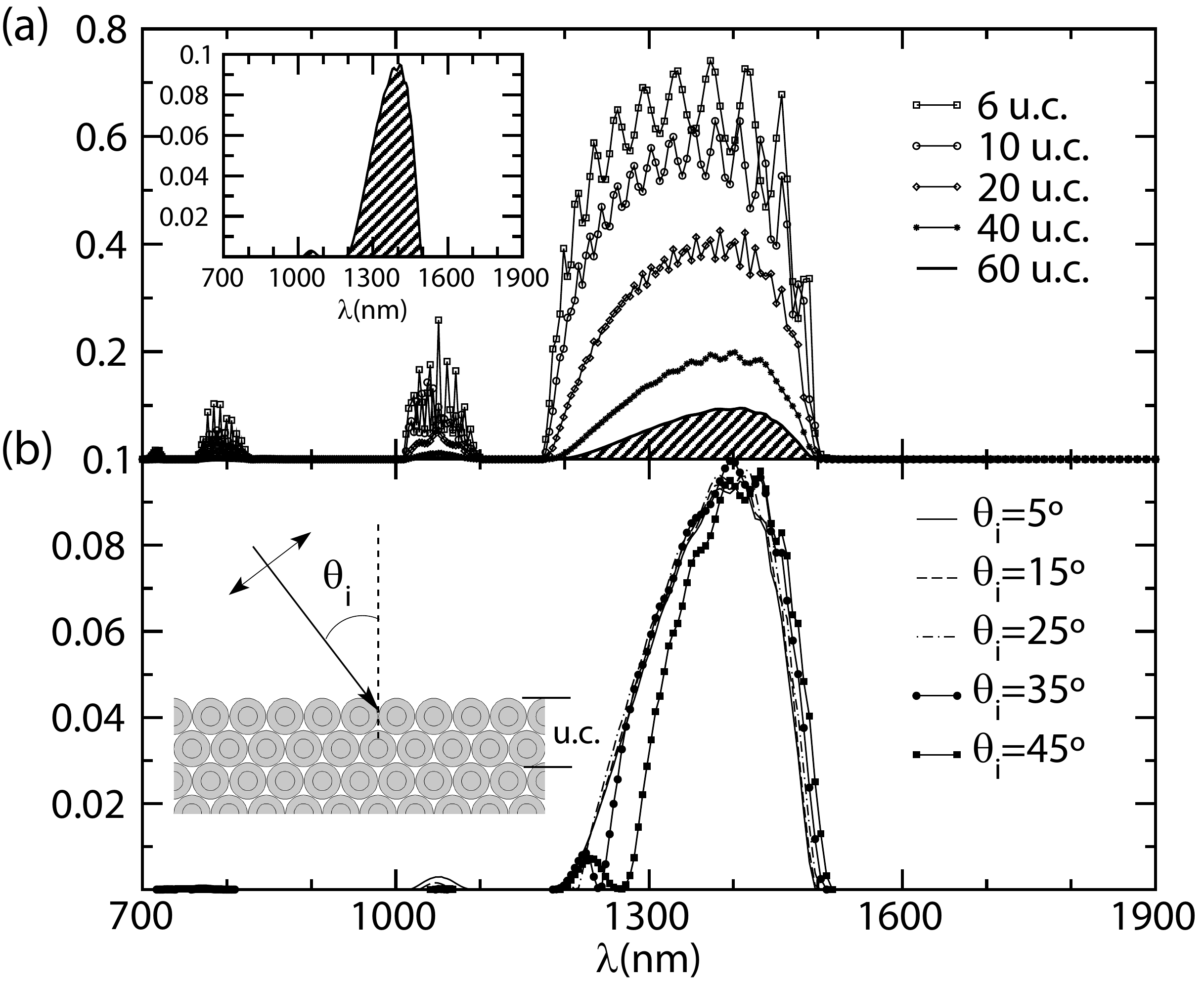}
\caption{(a) Transmission curves through metamaterial slabs made of a regular hexagonal arrangement of Ag@Si core-shell nanowire with external radius $R_{out}=170$~nm 
and internal radius $R_{int}=80$~nm. Center-to-center distance between neighbors is $L=350$~nm. Several
thicknesses under normal incidence illumination are considered, corresponding to different symbols. The inset shows the case of a slab with a 
thickness of 60 unit constituents (u.c.). (b) Transmission curves through a 60 u.c-thick slab under several angles of incidence.}
\label{figure5}
\end{figure}

First, we consider a periodic arrangement of Ag@Si core-shell nanocylinders in a hexagonal lattice. This configuration
is expected to lead to a highly isotropic effective response due to the high symmetry of the unit-cell (u-c). We take the geometry
of the meta-atoms to be the same as in figure \ref{figure4}. For these nanostructures there is a spectral overlap between the electric 
and magnetic resonances for $\lambda_0\in(1200,1500)$~nm. We expect a metamaterial with these
basic building blocks to present a left-handed (negative-index) frequency window in which propagation is allowed. To obtain 
this effective response, moderate filling fractions ($f$) are expected to be needed \cite{NJP}. Therefore, basic constituents are brought 
close together, fixing the surface-to-surface distance of every two neighboring nanoparticles to be $d_{s-s}=10$~nm. For a 
hexagonal lattice this implies a filling fraction of $f\sim0.856$. We have performed full numerical simulations using a Finite Element Method (FEM)
commercial software \cite{COMSOL}. Slabs are 
considered infinite along transverse direction by imposing appropriate periodic boundary conditions 
to the system. Let us now define a basic layer of these slabs to be such that different thicknesses are obtained by
adding a different number of layers, without need of any lateral translations. These basic layers are infinite in the transverse directions,
and we call them unit constituents of the slabs (u.c.).

\begin{figure}[!ht]
\includegraphics[width=\columnwidth]{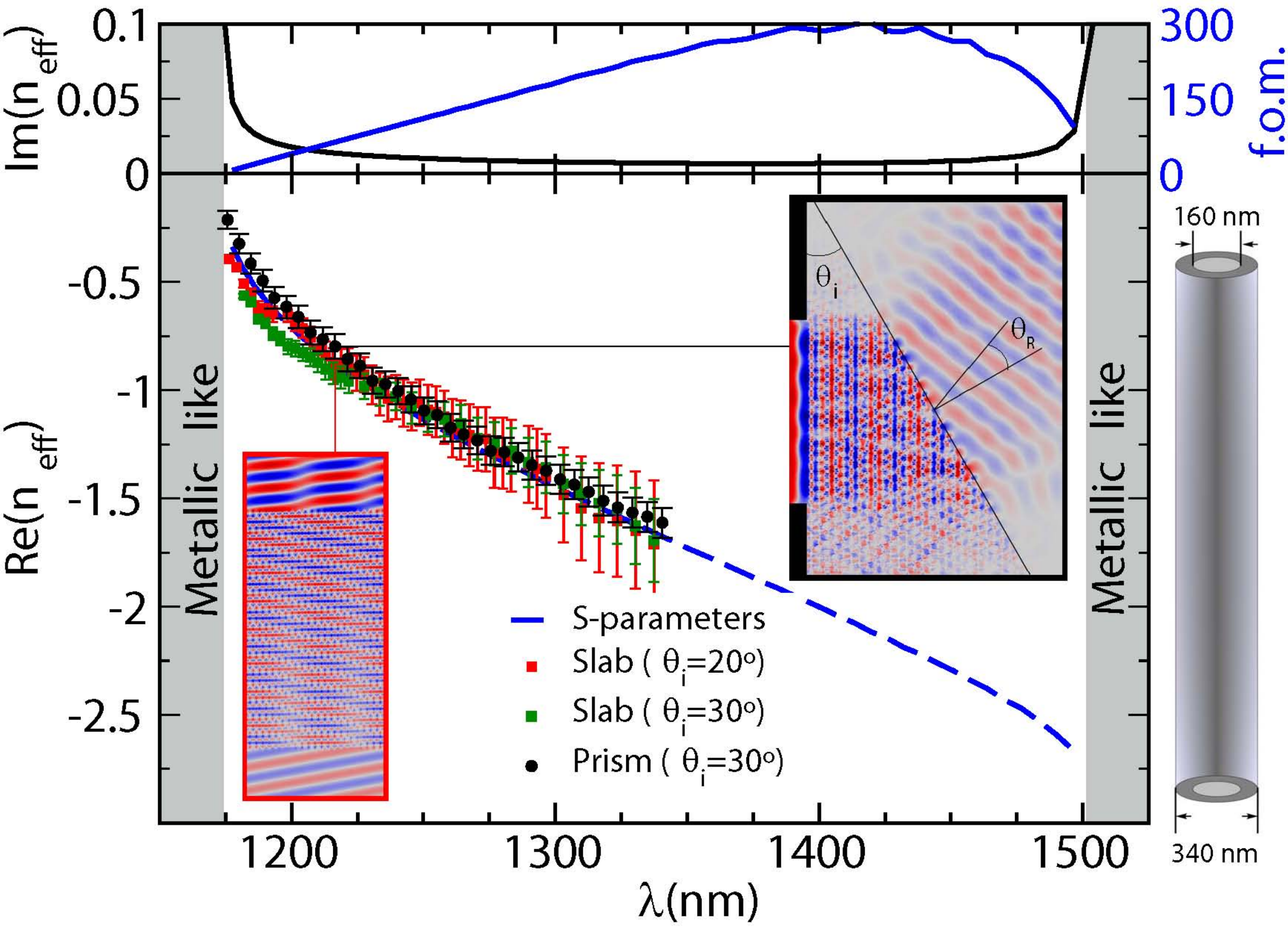}
\caption{Lower Panel: Retrieved real part of the index of refraction for a metamaterial built with Ag@Si core-shell nanowires with
$R_{int}=80$~nm and $R_{out}=170$~nm arranged in a hexagonal lattice. The blue discontinuous curve represents the values obtained 
from complex reflection and transmission coefficients under normal incidence.  Black circles represent the retrieved values 
from Snell's law in the prism configuration. The incidence angle is fixed, $\theta_{i}=\theta_{prism}\sim30\deg$.    
The corresponding inset shows the y-component of the electric field (only non-zero of the incident wave) 
for the particular case $\lambda\sim1215$~nm. Red and green squares represent the values obtained from Snell's law in the slab configuration. Angles of incidence
are $\theta_{i}=20\deg$ (red) and $\theta_{i}=30\deg$ (green). The corresponding inset represents the x-component (parallel to the first interface) of the 
electric field for $\lambda\sim1215$~nm. Upper Panel: Imaginary part of the index of refraction obtained from complex
reflection and transmission coefficients (black curve) and figure of merit, f.o.m., as defined in the text (blue curve).}
\label{figure6}
\end{figure}

In Fig.\ref{figure5}(a), the obtained transmission spectra under normal incidence illumination are depicted. Results
for slabs of several thicknesses, expressed in terms of the number of unit constituents, $N$, are included, the basic u.c. being described
in the inset of figure \ref{figure5}(b). A frequency window
in which propagation is allowed is present, which coincides with the range in which both resonances overlap. Thus, coupling 
between neighboring nanostructures does not seem to have a high impact, since the resonances are not critically modified. Recall
that, although $d_{s-s}$ is small compared to the size of the structures, the highest near field intensities are located in the
surface of the core. Since core-to-core distances, $d_{c-c}$, are of the order of $d_{c-c}\sim2(R_{out}-R_{int})$, coupling 
between plasmon modes is very weak, and the resonances are not disturbed. Modulations of the intensity inside the transmission 
window correspond to well known Fabry-Perot modes inside the slab, due to multiple reflections. The transmission peak arising at
higher energies ($\lambda\sim1050$~nm), slightly redshifted with respect to the electric-quadrupole-like resonance, almost completely disappears for 
sufficiently thick slabs, as evidenced in the 60 u.c.-thick slab case (see the inset in Fig.\ref{figure5}(a)). The obtained values 
within the transmission frequency range decrease with the thickness of the slab due to absorption in the system. Nevertheless, 
there is still a significant amount of energy transmitted through, even for such a thick slab as the one with 60 u.c. (i.e. for 
propagation lengths inside the NIM of about 21~$\mu$m). Within this frequency window, negative refraction is thus expected with low associated losses. Before retrieving the values of the effective parameters, let us show that 
this NIM is, moreover, highly isotropic. To do so, we have performed the same study for several angles of illumination, $\theta_i$. Results obtained for the 60 u.c.-thick 
slab and $\theta_i\in(0^{\circ},45^{\circ})$ are shown in Fig.\ref{figure5}(b). For low angles of incidence, 
$\theta_i\in(0^{\circ},35^{\circ})$, transmission curves are almost identical. For increasingly higher angles of incidence, 
$\theta_i\in(35^{\circ},45^{\circ})$ a Fabry-Perot modulation of the the transmission is observed in the high frequency limit. 

\begin{figure}[!ht]
\includegraphics[width=\columnwidth]{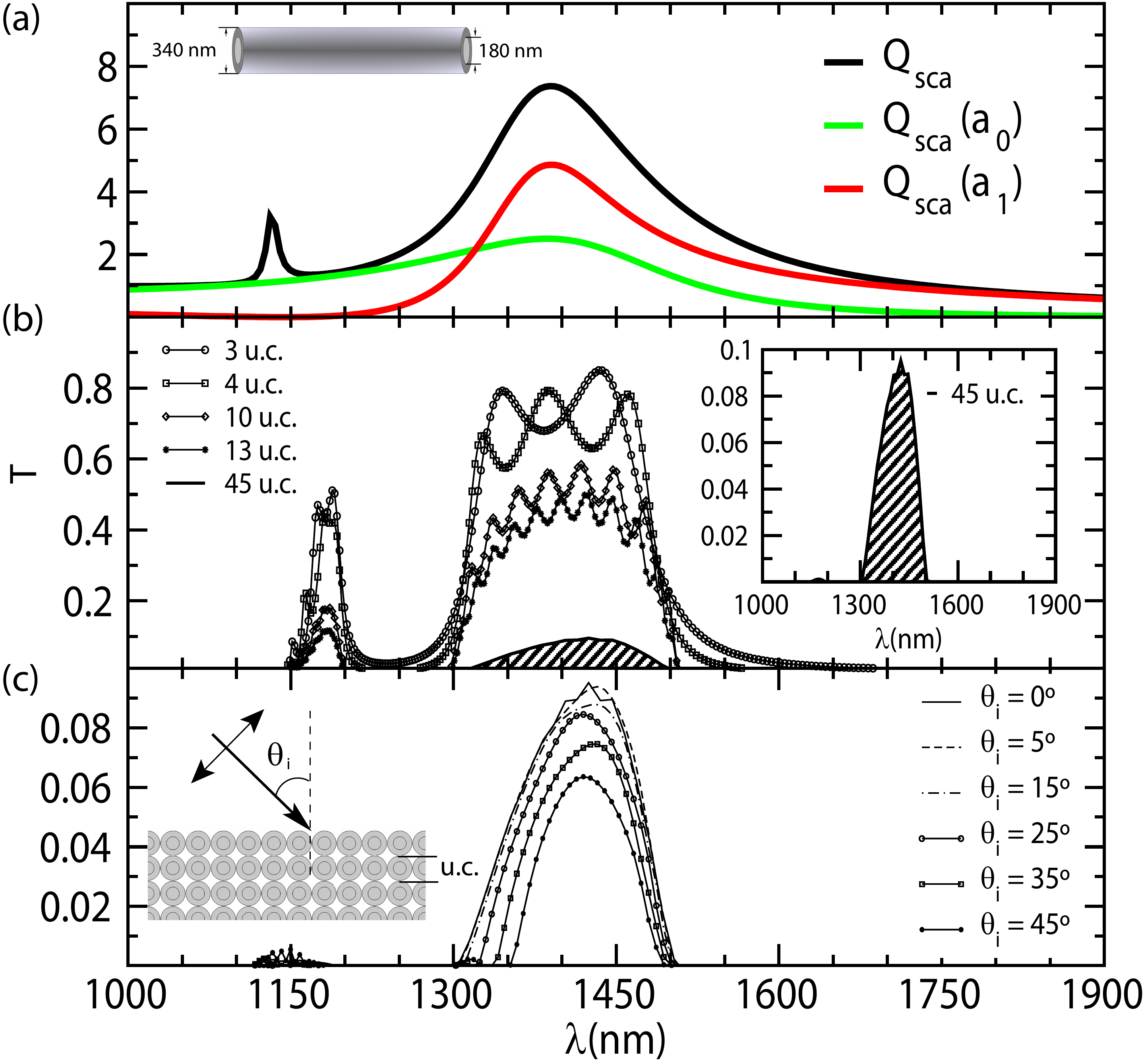}
\caption{(a) Scattering efficiency for a Ag@Ge core-shell nanowire with external radius $R_{out}=170$~nm and internal radius $R_{int}=90$~nm.
The contributions of the two first terms in the expansion of the field are also shown. (b) Transmission curves through metamaterial slabs of several
thicknesses under normal incidence illumination. The slabs are obtained by arranging these core-shell nanowires periodically in a square 
lattice (period $L=350$~nm). Curves with different symbols represent different slab thicknesses. The inset shows the case of a slab with a 
thickness of 45 unit constituents (u.c.). (c) Transmission curves through a 45 u.c-thick slab under several angles of incidence.}
\label{figure7}
\end{figure}

As previously mentioned, we took three alternative approaches to determine, within the transmission window, $Re(n_{eff})$ from simulations. Results obtained are 
depicted in the lower panel of Fig.\ref{figure6}. First, we have applied standard retrieval procedures \cite{RetPar,RobRetPar}, which allows one to determine the effective 
parameters from complex reflection and transmission coefficients (or S-parameters) under normal incidence illumination, assuming an equivalent homogeneous 
slab. The blue dashed curve represents the obtained values. We have then applied Snell's law to those slab configurations in which the angle of incidence is 
$\theta_i\neq0$. The results for the particular case of a 20 u.c.-thick slab under $\theta_i=20^{\circ},30^{\circ}$ illumination are plotted as red and green squares. Additionally, we 
simulated a rectangular prism with an angle $\theta_{prism}\sim30^{\circ}$ and applied Snell's law: the thus retrieved values are indicated by black circles. Both 
insets represent the electric field component parallel to the first interface in the slab and prism configurations for $\lambda\sim1215$~nm.
Negative refraction with extremely low losses is clearly apparent. Moreover, phase reversal, characteristic of systems having 
simultaneously $\epsilon<0$ and $\mu<0$, can be observed plotting dynamically this component in a full harmonic cycle (not shown here).  Errors in those values
retrieved graphically from Snell's law, have been computed by assuming a systematic error in the determination of the angle of $\pm2^{\circ}$.
It is apparent from Fig.\ref{figure6} that the retrieved values are consistent regardless of the angle of incidence. Therefore, as expected, the obtained
metamaterial is highly isotropic.

\begin{figure}[!ht]
\includegraphics[width=\columnwidth]{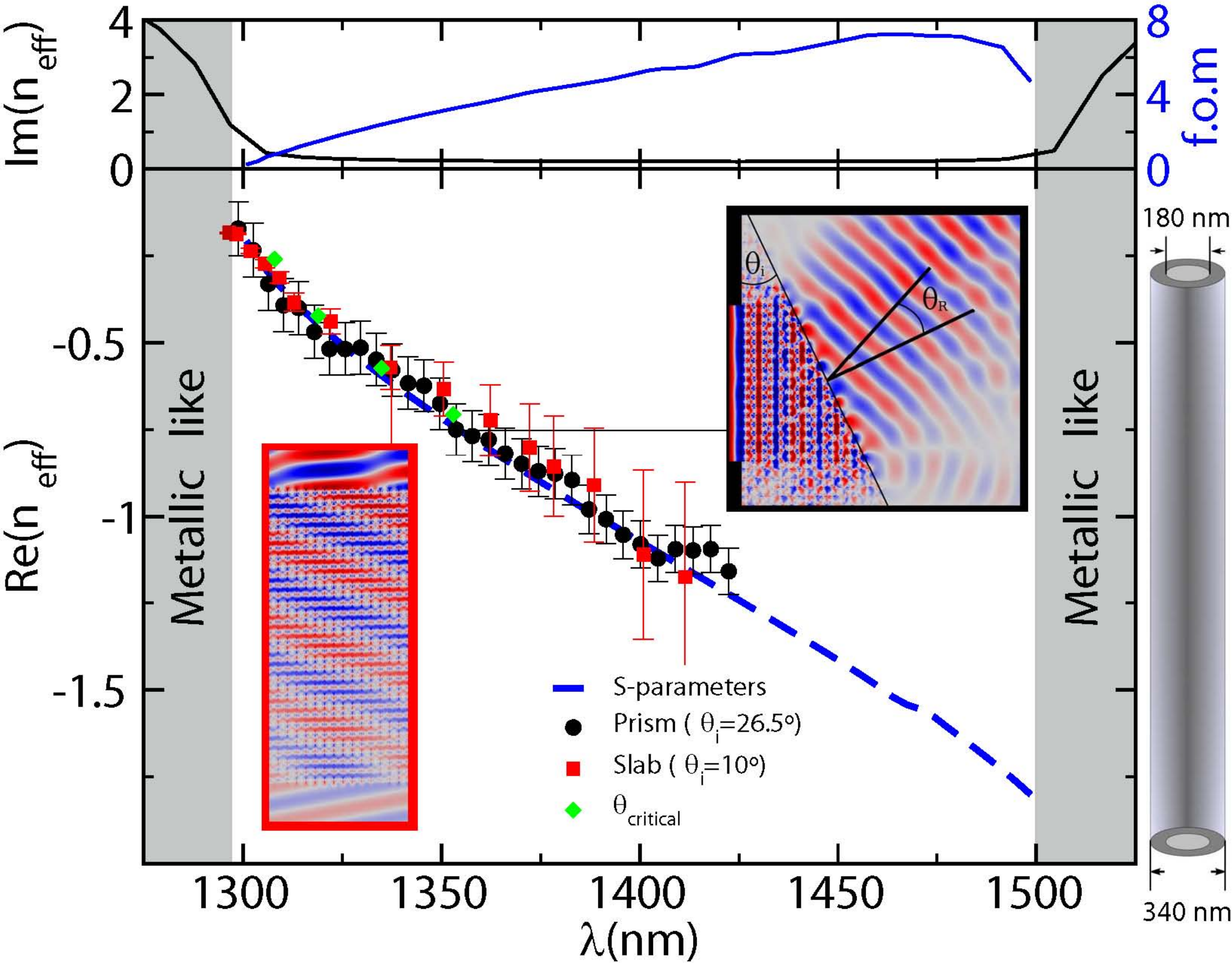}
\caption{Lower Panel: Retrieved real part of the index of refraction using Snell's law for a metamaterial built with Ag@Ge core-shell nanowires with
$R_{int}=90$~nm and $R_{out}=170$~nm arranged in a square lattice. The blue discontinuous curve represents the values obtained from complex reflection and transmission coefficients of a 45 u.c.-thick slab under normal incidence.  Black circles represent the retrieved values from Snell's law in
the prism configuration. The incidence angle is fixed, $\theta_{i}=\theta_{prism}\sim26.5\deg$.    
The corresponding inset shows the y-component of the electric field (only non-zero of the incident wave) 
for the particular case $\lambda\sim1355$~nm. Red squares represent the values obtained from Snell's law in the slab configuration. The angle of incidence
is $\theta_{i}=10\deg$. The corresponding inset represents the x-component (parallel to the first interface) of the electric field for $\lambda\sim1335$~nm. Green diamonds are the values retrieved from critical angles for total internal reflection.
Upper Panel: Imaginary part of the index of refraction obtained from complex
reflection and transmission coefficients (black curve) and figure of merit, f.o.m., as defined in the text (blue curve)}
\label{figure8}
\end{figure}

It is important to mention that, although the standard retrieval procedure based on S-parameters allows one to compute the index of refraction in the whole frequency range, the effective 
medium description of the system breaks down at longer wavelengths inside the transmission window. If somehow counterintuitive, this effect is a consequence 
of the fact that $\partial Re(n_{eff}(\omega))/\partial\omega>0$, leading to higher absolute values at longer wavelengths, which result in shorter 
effective wavelengths inside the metamaterial, $\lambda_{eff}$. In this way, at $\lambda_0=1150$~nm, the effective wavelength inside the structure 
can reach $\lambda_{eff}=\lambda_0/|Re(n_{eff})|\sim15$~$\mu$m, i.e., up to 40 times the characteristic period length within the structure. On the 
other hand, at $\lambda_0=1500$~nm, e.g., $\lambda_{eff}=\lambda_0/|Re(n_{eff})|\sim575$~nm. This implies a wavelength inside the system of less than 
two times the period. Consequently, the effective medium fails to properly describe the metamaterial properties.

With respect to the losses, apart from the inset contour maps, there is a strong evidence that the metamaterial is very weakly absorbing, as deduced from the fact
that the transmission values are as high as 0.1 within the negative-index band, with
about the same portion of reflection, for propagation lengths inside the metamaterial of about 21~$\mu$m. One can
actually compute the expected values of the imaginary part of the index through the standard S-parameter retrieval procedure: The results are plotted in the upper panel
of figure \ref{figure6}. While values as extraordinarily low as $7\times10^{-3}$ are achieved within the transmission window, out of it, the metamaterial
essentially behaves as a metal, with $Re(n_{eff})<Im(n_{eff})$. Inside the negative-index band, values of the figure of merit (f.o.m$=-Re(n_{eff})/Im(n_{eff})$) can, in turn, be as high as f.o.m.$\sim300$. In the particular case of $Re(n_{eff})=-1$,  f.o.m.$\sim85$, i. e., $Im(n_{eff})=0.012$.

Let us next study the case of both different arrangement and semiconductor shell: a square lattice of Ag@Ge core-shell nanowires. Since the square lattice possesses lower symmetry than the hexagonal, it would not be surprising to find 
some degree of anisotropy. In Fig.\ref{figure7}(a) we have plotted the optical 
properties of a Ag@Ge core-shell nanowire with external radius $R_{out}=170$~nm and internal radius $R_{int}=90$~nm. For this nanostructure there is a 
spectral overlap between the electric and magnetic resonances for $\lambda_0\in(1300,1500)$~nm. We have also performed full numerical simulations for this system,
considering several slabs with different thicknesses obtained by arranging these nanowires in a simple square lattice with lattice period $L=350$~nm, 
thus keeping fixed the surface-to-surface distance between neighboring particles ($d_{s-s}=10$~nm): this leads to $f\sim0.74$. 
Therefore, we expect to achieve lower absolute values of the real part of the index. 

In Figure~\ref{figure7}(b), the obtained 
transmission spectra under normal incidence illumination are depicted. The frequency window in which propagation is allowed 
also presents Fabry-Perot modes. For the square lattice arrangement, typically, $N-1$ peaks are present 
along the propagation direction. The transmission peak arising at higher energies ($\lambda\sim1175$~nm), also disappears for sufficiently thick slabs, as 
evidenced in the 45 u.c.-thick slab case (see the inset in Fig.\ref{figure7}(b)). Again, we expect the refractive index of the metamaterial
to be negative in the transmission window, with low associated losses, although slightly higher than in the previous Ag@Si case (notice that the same transmission is present in the $N=45$ case, which
implies a propagation length inside the NIM of about 16~$\mu$m). Similarly to the Ag@Si case, several angles of  illumination, $\theta_i$ have been considered. The results obtained for the 45 u.c.-thick slab and $\theta_i\in(0^{\circ},45^{\circ})$ are shown in Fig.\ref{figure7}(c). For low 
angles of incidence, $\theta_i\in(0^{\circ},25^{\circ})$, transmission curves are, again, almost identical. However, in this case for high enough angles 
of incidence, $\theta_i\in(25^{\circ},45^{\circ})$ there is a narrowing of the transmission window in the high frequency limit. As 
will be shown later, in that region the real part of the effective index of refraction ($n_{eff}$) reaches its lowest values (even approaching zero, 
$|Re(n_{eff})|\sim0$). For those wavelengths for which the transmission suddenly drops, the incidence angles are above corresponding critical angles for 
total internal reflection. Therefore, no transmission is allowed. This allows for an alternative way to determine the real part of the index. 

Finally, the results for $Re(n_{eff})$ obtained from simulations within the transmission window are depicted in the lower panel of Fig.\ref{figure8}. The blue dashed curve represents 
the obtained values from S-parameter retrieval procedures. Red squares are the obtained values from Snell's law applied to a 30 u.c.-thick slab 
under $\theta_i=10^{\circ}$ illumination. Black circles are the values retrieved from the simulation of a rectangular prism with an angle 
$\theta_{prism}\sim26.5^{\circ}$. Additionally, the transmission curves under oblique incidence, in which the incident wave experiences total internal 
reflection, enables to obtain $Re(n_{eff})$ at the wavelength for which $T=0$, by simply applying $Re(n_{eff})=-sin(\theta_i)$ (green diamonds 
in Fig.\ref{figure8}). Again, the retrieved values for $Re(n_{eff})$ seem to be consistent regardless of the angle of incidence, even though 
one could expect the system to present  anisotropy. The error bars in those results stemming from the application of Snell's law have been
estimated assuming, again, a systematic error of $\pm2^{\circ}$ in the determination of the angle. As expected, absolute values of the index of refraction are
lower in this configuration, due to the lower value of the filling fraction, but NIM behavior with $n_{eff}\sim -1$ is clearly observed too.
Although higher in this case (see upper panel of figure \ref{figure8}), losses are still very low. At $\lambda\sim1390$~nm, for which
$Re(n_{eff})=-1$, the f.o.m. reaches 4.7, implying an imaginary part of the index of only $Im(n_{eff})=0.21$.

In conclusion, as showed here, metallo-dielectric core-shell nanowires of circular cross section, are good candidates as building blocks of highly 
isotropic, negative index metamaterials operating at optical frequencies, with extremely low losses exhibiting figures of merit f.o.m.$\sim 300$ (two orders of magnitude larger than previously proposed NIMs). Although here demonstrated to operate at 
$\lambda_0\in(1.15,1.35)$~$\mu$m for Ag@Si with $R_{in}=80$~nm  and $R_{out}=170$~nm arranged in a hexagonal lattice, and at
$\lambda_0\in(1.3,1.45)$~$\mu$m for Ag@Ge nanowires with $R_{in}=90$~nm and $R_{out}=170$~nm arranged in a square lattice, negative refraction in the whole telecom frequency range can be achieved by tuning the geometrical 
parameters. Moreover, the operating physical principles can be applied 
to obtain negative refraction in completely different ranges of the electromagnetic spectrum (IT and THz) by using appropriate materials.

The authors acknowledge M. Nieto-Vesperinas and R. Marqu{\'e}s for fruitful discussion in the preparation of this manuscript. Also acknowledge
the Spain Ministerio de Econom{\'{i}}a y Competitividad, through the
Consolider-Ingenio project EMET (CSD2008-00066) and NANOPLAS (FIS2009-11264), and
the Comunidad de Madrid (grant MICROSERES P2009/TIC-1476) for support. R.~P.-D., also acknowledges support from The European Social Fund and CSIC through a JAE-Pre grant.


\begin{thebibliography}{}
\bibitem{PendryPRL} J. B. Pendry, {\itshape{Phys. Rev. Lett.}} {\bf{2000}}, 85, 3966.
\bibitem{NatMat} X. Zhang and Z. Liu, {\itshape{Nat. Materials}} {\bf{2008}}, 7, 435.
\bibitem{HyperMet} Z. Jacob, I. I. Smolyaninov and E. E. Narimanov, {\itshape{Appl. Phys. Lett.}} {\bf{2012}}, 100, 181105.
\bibitem{TO} J. B. Pendry, D. Schurig, and D. R. Smith, {\itshape{Science}} {\bf{2006}}, 312, 1780.
\bibitem{TO2} U. Leonhardt, {\itshape{Science}} {\bf{2006}}, 312, 1777.
\bibitem{OptMet} V. M. Shalaev, {\itshape{Nat. Photonics}} {\bf{2007}}, 1, 41.
\bibitem{WegSou} C. M. Soukoulis and M. Wegener, {\itshape{Nat. Photonics}} {\bf{2011}}, 154, 1.
\bibitem{PRLSat} J. Zhou, Th. Koschny, M. Kafesaki, E. N. Economou, J. B. Pendry and C. M. Soukoulis, {\itshape{Phys. Rev. Lett.}} {\bf{2005}}, 95, 223902.
\bibitem{PRLfish} Carlos García-Meca, Juan Hurtado, Javier Martí, Alejandro Martínez, Wayne Dickson, and Anatoly Zayats,  {\itshape{Phys. Rev. Lett.}}  \textbf{106}, 067402 (2011).
\bibitem{Engheta} A. Al\'u, A. Salandrino and N. Engheta, {\itshape{Opt. Express}} {\bf2006}, 14, 1557.
\bibitem{Marques} L. Jelinek and R. Marqu\'es, {\itshape{J. Phys.: Condens. Matter}} {\bf2010}, 22, 025902.
\bibitem{Wheeler} M. S. Wheeler, J. S. Aitchison, and M. Mojahedi, {\itshape{Phys. Rev. B}} {\bf2005}, 72, 193103.
\bibitem{PRLPopa} B. I. Popa and S. A. Cummer  {\itshape{Phys. Rev. Lett.}} {\bf{2008}}, 100, 207401.
\bibitem{G-E} A. Garc\'ia-Etxarri, R. G\'omez-Medina, L. S. Froufe-P\'erez, C. L\'opez, L. Chantada, F. Scheffold, J. Aizpurua, M. Nieto-Vesperinas, and J. J. S\'aenz, {\itshape{Opt. Express}} {\bf2011}, 19, 4815.
\bibitem{MatTod} Q. Zhao, J. Zhou, F. Zhang and D. Lippens, {\itshape{Mat. Today}} {\bf{2009}}, 12, 60.
\bibitem{PRLDie} K. Vynck, D. Felbacq, E. Centeno, A. I. C{\^{a}}buz, D. Cassagne and B. Guizal, {\itshape{Phys. Rev. Lett.}} {\bf{2009}}, 102, 133901.
\bibitem{PRLDie2} J.A. Schuller, R. Zia, T. Taubner and M. L. Brongersma, {\itshape{Phys. Rev. Lett.}} {\bf{2007}}, 99, 107401.
\bibitem{Kerker} M. Kerker and E. Matijevi{\'{c}}, {\itshape{J. Opt. Soc. Am.}} {\bf{1961}}, 51, 506.
\bibitem{NJP} R. Paniagua-Dom{\'{i}}nguez, F. L\'opez-Tejeira, R. Marqu\'es, and J. A. S\'anchez-Gil, {\itshape{New J. Phys.}} {\bf{2011}}, 13, 123017.
\bibitem{COMSOL} COMSOL Multiphysics {\bf{2012}} version 4.2.
\bibitem{RetPar} Smith, D. R., Vier, D. C., Koschny, T. and Soukoulis, C. M. {\itshape{Phys. Rev. E}} {\bf{2005}}, 71, 036617.
\bibitem{RobRetPar} Chen, X., Grzegorczyk, T. M., Wu, B., Pacheco, J. Jr. and Kong J. A.  {\itshape{Phys. Rev. E}} {\bf{2004}}, 70, 016608. 
\end{thebibliography}
\end{document}